\documentclass{article}
\usepackage{spconf}
\usepackage{amsmath,mathtools,bbm}
\usepackage{amsfonts}
\usepackage{graphicx}
\usepackage[export]{adjustbox} % valign=m for tabular
\usepackage{caption}
\usepackage{subcaption}
\usepackage{multirow}
\usepackage{booktabs} % pour faire des \cmidrule(lr){2-4}
\usepackage[colorlinks=true]{hyperref} % pagebackref=true

\usepackage{lipsum}

\graphicspath{{./pictures/}}

\usepackage{xcolor}
\newcommand{\juju}[1]{\textcolor{blue}{J:\bf [#1]}}

\newcommand{\new}[1]{\textcolor{black}{#1}}
\newcommand{\old}[1]{\textcolor{red}{}}

\newlength{\imgsize}

\title{
    {L{\small ATENT}P{\small ATCH}: A NON-PARAMETRIC APPROACH FOR FACE GENERATION AND EDITING}
}
\name{Benjamin Samuth \qquad Julien Rabin \qquad David Tschumperlé \qquad Frédéric Jurie}
\address{Normandie Univ., UNICAEN, ENSICAEN, CNRS, GREYC, 14000 Caen, France
\\
\small\texttt{\{Benjamin.Samuth, Julien.Rabin,  David.Tschumperle, Frederic.Jurie\}@unicaen.fr}
}
\begin{document}
\twocolumn[{%
  \renewcommand\twocolumn[1][]{#1}%
  \maketitle % \maketitle DANS \twocolmun
  % ---- TEASER -----
  \begin{center}
    \centering
    \setlength\tabcolsep{0.7mm}
    \setlength\imgsize{0.155\textwidth}
    \newcommand{\teasersubcaption}[3]{\parbox{#1}{\vspace{0.4em}\centering\small #2 \newline \footnotesize #3}}
    %\begin{tabular}{c cccccc}
    %{c p{\imgsize}>{\centering} p{\imgsize}>{\centering} p{\imgsize}>{\centering} p{\imgsize}>{\centering} p{\imgsize}>{\centering} p{\imgsize}>{\centering}}
    \begin{tabular}{c p{\imgsize} p{\imgsize} p{\imgsize} p{\imgsize} p{\imgsize} p{\imgsize}}
        \rotatebox[origin=c]{90}{\small Data}        &\multicolumn{6}{c}{\includegraphics[valign=m,width=.96\linewidth]{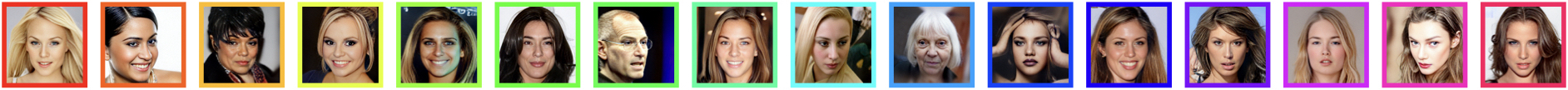}}
        \\[4mm]
        \rotatebox[origin=c]{90}{\small Generated -- Code}
        %&\multicolumn{6}{c}{\includegraphics[valign=m,width=.95\linewidth]{pictures/teaser/newteaser/gen_image_with_codes_V2.jpg}}
           &\includegraphics[valign=m,width=\imgsize]{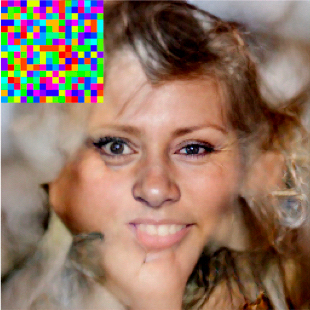} 
           &\includegraphics[valign=m,width=\imgsize]{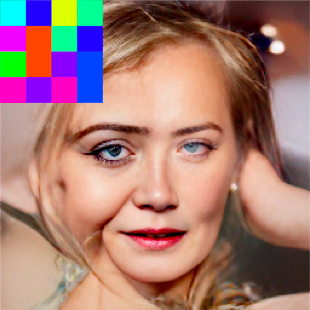}  
           &\includegraphics[valign=m,width=\imgsize]{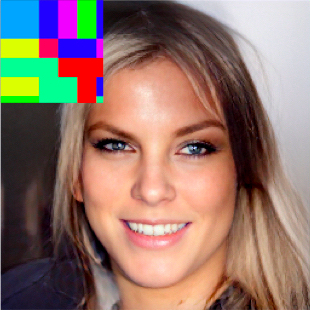} 
           &\includegraphics[valign=m,width=\imgsize]{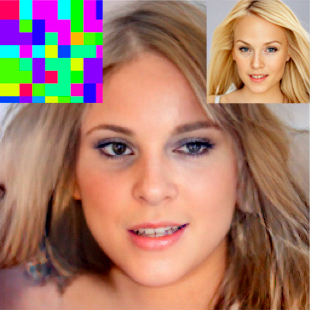} 
           &\includegraphics[valign=m,width=\imgsize]{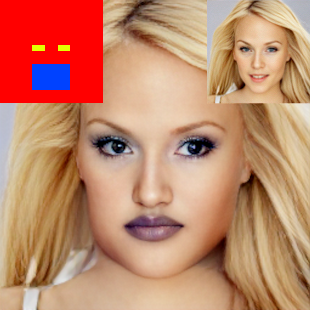} 
           &\includegraphics[valign=m,width=\imgsize]{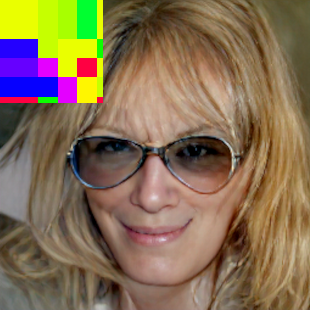} 
          \\
          & \teasersubcaption{\imgsize}{(a) Random Code}{1×1 Patch copy}
          & \teasersubcaption{\imgsize}{(b) Random Code}{4×4 Patch copy}
          & \teasersubcaption{\imgsize}{(c) {\sc LatentPatch}}{2×2 Patch sequence}
          & \teasersubcaption{\imgsize}{(d) Using reference}{First image, in red}
          & \teasersubcaption{\imgsize}{(e) Face edition}{Eyes and mouth}
          & \teasersubcaption{\imgsize}{(f) Selected attributes}{Woman, Blond, Glasses}
          \\[-2mm]
    \end{tabular}
   
    % \begin{tabular}{ccc|c|c}
    %   \rotatebox[origin=c]{0}{\small Ground Truth}
    %   &\rotatebox[origin=c]{0}{\small 1×1 Random}
    %   &\rotatebox[origin=c]{0}{\small Explainability}
    %   &\rotatebox[origin=c]{0}{\small Reference-based}
    %   \\
    %   \includegraphics[valign=m,width=\imgsize]{teaser/gt.png}
    %   &\includegraphics[valign=m,width=\imgsize]{teaser/1x1.png}
    %   &\includegraphics[valign=m,width=\imgsize]{teaser/gen_id.png}
    %   &\includegraphics[valign=m,width=\imgsize]{teaser/by_ref.png}
    %   &\includegraphics[valign=m,width=\imgsize]{teaser/edit_id.png}        
    %   \\
    %   \includegraphics[valign=m,width=\imgsize]{teaser/rec.png}
    %   &\includegraphics[valign=m,width=\imgsize]{teaser/4x4.png}
    %   &\includegraphics[valign=m,width=\imgsize]{teaser/gen.png}
    %   &\includegraphics[valign=m,width=\imgsize]{teaser/by_attr.png}
    %   &\includegraphics[valign=m,width=\imgsize]{teaser/edit_result.png}
    %   \\
    %   \rotatebox[origin=c]{0}{\small Reconstruction}
    %   &\rotatebox[origin=c]{0}{\small 4×4 Random}
    %   &\rotatebox[origin=c]{0}{\small Our method (16 images)}
    %   &\rotatebox[origin=c]{0}{\small Attribute-constrained}
    %   &\rotatebox[origin=c]{0}{\small Editing}
    % \end{tabular}
    
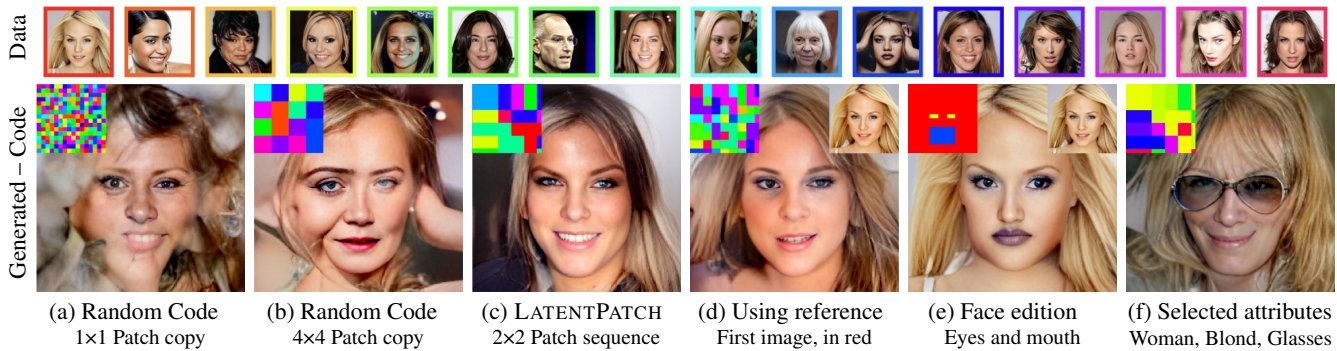
\captionof{figure}{\small The proposed patch-based approach, coined ‘‘{\sc LatentPatch}’’, can generate images like (c) using only the 16 source images shown in the first row, without any learning. It also enables easy implementation of variants such as (d) reference-based generation, (e) editing, and (f) attribute-constrained generation (on alternative data not shown here). Images (a) and (b) were generated using random patches from the source images, with the origin of each patch indicated by its color. Additional results can be found on the project page~\cite{samuth2023webpage}.
    %Our approach presents a simple solution for few-shot face generation utilizing non-parametric synthesis algorithms within the latent space of an auto-encoder.  The first row shows the source dataset of 16 images (with color codes) randomly sampled from CelebA-HQ and reconstructed with VQ-GAN~\cite{esser2021taming} pre-trained on FFHQ. We demonstrate that the decoder of the VQ-GAN can proficiently extrapolate a wide range of spatial arrangements of facial latent patches (a \& b). By merging its capability to abstract local features with non-parametric generation techniques, we are then able to generate new samples from a small number of real images while maintaining explicit explainability of the data used (c).  This approach allows for straightforward implementation of multiple variants, such as (d) referenced and (e) editing or (f) attribute-constrained generation, without requiring additional fine-tuning. More results are available on the project page~\cite{samuth2023webpage}.
    }
  \label{fig:teaser}
  \end{center}

}]

% Je trouve l'ancien abstract trop long et détaillé (redondance avec l'intro) ce qui n'est pas top pour un papier de 4 pages
% l'ancienne version est dispo dans le fichier "old_stuff removed by fj.tex"
\begin{abstract} This paper presents LatentPatch, a new method for generating realistic images from a small dataset of only a few images. We use a lightweight model with only a few thousand parameters. Unlike traditional few-shot generation methods that fine-tune pre-trained large-scale generative models, our approach is computed directly on the latent distribution by sequential feature matching, and is explainable by design. Avoiding large models based on transformers, recursive networks, or self-attention, which are not suitable for small datasets, our method is inspired by non-parametric texture synthesis and style transfer models, and ensures that generated image features are sampled from the source distribution. We extend previous single-image models to work with a few images and demonstrate that our method can generate realistic images, as well as enable conditional sampling and image editing. We conduct experiments on face datasets and show that our simplistic model is effective and versatile.
\end{abstract}


\begin{figure*}[!htb]
    \centering
    \includegraphics[width = 0.96\linewidth]{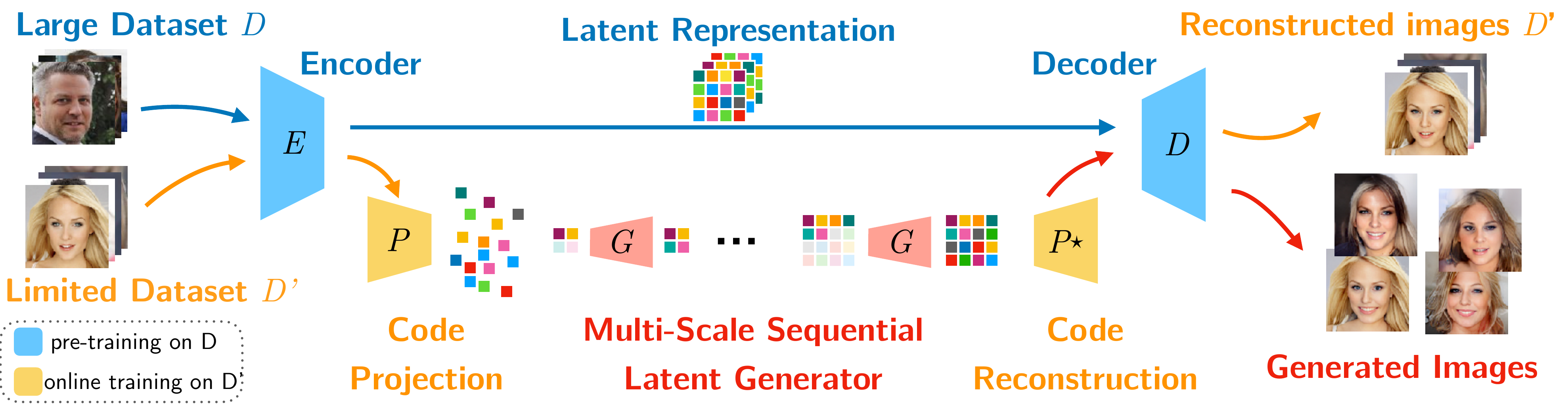}
    \caption{\small 
        Illustration of the proposed face generation framework based on a pre-trained auto-encoder.
        See the text for more details.
        %An auto-encoder is trained on a large dataset and latent codes from a limited dataset are used to generate original images.
    }
    \label{fig:schema}
    \vspace{-1em}
\end{figure*}

\section{Introduction}
\label{sec:intro}
% # realistic image generation
% Style-GAN~\cite{karras2019style}
% VQ-VAE2~\cite{razavi2019generating}
% VQ-GAN~\cite{esser2021taming}

% # conditional
% BIG-GAN

% # text 2 image models:
% GLIDE~\cite{nichol2021glide}
% DALL-E~\cite{ramesh2022hierarchical}
% IMAGEN~\cite{saharia2022photorealistic}
% latent diffusion model~\cite{rombach2022high}

Deep generative networks have made significant progress in photo-realistic image synthesis by using adversarial or diffusion models and training on large-scale datasets. Currently, the most advanced text-to-image generation architectures, such as 
%GLIDE~\cite{nichol2021glide}, %DALL-E~\cite{ramesh2022hierarchical}, %IMAGEN~\cite{saharia2022photorealistic}, and 
latent diffusion~\cite{rombach2022high}, 
are comprised of billions of trainable parameters, requiring datasets of similar magnitude for training (\emph{e.g.} LAION-5B). Recent efforts have focused on training generative models from small datasets, including those composed of just a few or even a single image. However, to produce realistic samples that are indistinguishable from true examples by human evaluation, even within a more specific image domain, such as human faces~\cite{karras2019style} or indoor scenes, %~\cite{yu2015lsun},
generative models require deep and wide neural networks (\emph{e.g.} StyleGAN~\cite{karras2019style}, VQ-VAE~\cite{razavi2019generating} or VQ-GAN~\cite{esser2021taming}). 
In some cases, building large datasets is not possible, such as in medical imaging where data is scarce,  and data augmentation techniques may not be applicable~\cite{karras2020training}. 
Moreover, even with a large dataset, there is no guarantee that a large capacity model 
will not memorize some of the training samples~\cite{webster2019detecting,webster2021person,esser2021taming}. %
This issue has recently raised concerns about confidentiality, privacy, and copyright~\cite{somepalli2022diffusion,carlini2023extracting}.
Additionally, training large models requires a significant amount of computational resources
(in terms of \emph{RAM} and \emph{GPU-days}~\cite{rombach2022high}),
%%, ranging from tens to thousands of \emph{GPU-days}~\cite{rombach2022high}, and evaluation requires large memory requirements, 
%%which may be excessive for some applications.
which seems excessive for some applications.
 
Given the difficulty of training models on very small image sets, different techniques have been employed. One approach, known as \emph{few-shot generation}, is based on knowledge distillation, which involves fine-tuning a large model to a small dataset. This approach has been widely used for GANs, as evidenced by FreezeD~\cite{mo2020freeze}, TGAN~\cite{wang2018transferring}, MineGAN~\cite{wang2020minegan}, FS-GAN~\cite{robb2020few}, and~\cite{zhao2022closer}. It has also been successfully applied to diffusion models, including text-to-image models, such as in the case of~\cite{zhu2022few}.
%and~\cite{ruiz2022dreambooth}. \juju{[verif pertinence]}

% \juju{citer d'abord}
% few thousand images (or 100 of the same person)
% - training deep generative model with limited data~\cite{karras2020training}
% - \juju{référence recente intéressante : \cite{zhong2022deep}}
% \juju{puis}
% - FewGAN~\cite{ben2022fewgan} train an autoregressive model (pixel CNN / patch GAN) on a arbitrarily small dataset (ranging from 1 to 10)
% \juju{nombre de paramètres ?}

%\cite{karras2020training, zhao2020differentiable, zhong2022deep}
An alternative approach is to use differentiable data augmentation methods~\cite{karras2020training, zhao2020differentiable} to train the models from scratch. This allowed for the training of GANs on a more diverse dataset, which can be reduced to just a few thousand face images. Even fewer images may be used if they are perceptually similar, such as multiple images of the same person. As far as we know, FewGAN~\cite{ben2022fewgan} is the only autoregressive generative model that has been trained on a very small dataset of landscapes. %\juju{à compléter ?} 

One extreme example is the case of single image generation, as proposed by SinGAN~\cite{shaham2019singan}.
%\cite{shaham2019singan, ulyanov2016texture, bergmann2017learning, leclaire2021stochastic, houdard2022generative}
Texture synthesis is a related proxy problem, for which various models have been proposed~\cite{shaham2019singan, ulyanov2016texture, leclaire2021stochastic, houdard2022generative}. These last two applications can be viewed as a type of image reshuffling, which can be achieved through patch-based sampling techniques, such as those described in~\cite{efros1999texture}. Recently, GPNN~\cite{granot2022drop} and PSIN~\cite{cherel2022patch} have demonstrated that generative models are not always necessary to synthesize high-quality random samples. This can be achieved using a variant of the Patch-Match algorithm~\cite{barnes2010generalized} with GPU-based acceleration. Patch sampling has also been extended to latent representations in recent applications to image stylization~\cite{samuth2022patch} and inpainting~\cite{cherel2022attention}.

%In this work, we investigate the extension of these non-parametric techniques for lightweight image generation from an arbitrarily small dataset. This is a difficult problem on which  no patch-based techniques have been successful yet. For instance \cite{ben2022fewgan} showed that using approached such as~\cite{shaham2019singan,granot2022drop} on the concatenated set of images produces an irrelevant ‘‘patchwork’’.  

Our method is positioned at the intersection of the three types of aforementioned approaches: generative methods based on patches, from-scratch training, and adaptation of pre-trained models. What makes our method effective even in the presence of an extremely limited number of images is that we mix three key ingredients. Firstly, we rely on an auto-encoder that is pre-trained on a large image dataset. We use this auto-encoder as a tool to produce a universal latent compact representation (the encoder) capable of decoding such representation into images (the decoder). 
\old{Secondly, we adapt to the manifold described by the source image set simply through principal component analysis (PCA), which is simpler than finetuning or adapting a large model as done by other methods like~\cite{mo2020freeze,wang2018transferring,wang2020minegan}.}
\new{Secondly, we compress the manifold described by the source image set simply through principal component analysis (PCA), in order to improve the computation time and the memory usage of our method.}
Finally, our generative model is a simple patch-based model that does not require any training, in contrast with high-capacity latent generative models (such as transformers in~\cite{esser2021taming}, auto-regressive pixelCNN~\cite{van2016conditional} in~\cite{razavi2019generating, ben2022fewgan}, or cross-attention layers in~\cite{rombach2022high}). Our model resorts to multi-scale latent patch combinations to generate plausible latent code sequences from the source dataset, which is a non-parametric approach \new{of the generative part, as opposed to finetuning or adapting larger models as done by other methods like~\cite{mo2020freeze,wang2018transferring,wang2020minegan}}. %The proposed approach is illustrated in Fig.~\ref{fig:teaser} and summarized in Fig.~\ref{fig:schema}. Our experiments in Section~\ref{sec:experiments} demonstrate the effectiveness of this simple model even for limited source datasets. We show that it is capable of quickly generating and editing images in a very realistic way.

%\juju{citer MUSE \cite{chang2023muse} ? et \cite{preechakul2022diffusion} pour les AE + diffusion ? }

% = = = = = = = = = = = = = = = = = = = = = = = = = = = = = = = = = = = =
%  = = = = = = = = = = = = = = = = = = = = = = = = = = = = = = = = = = = =
% = = = = = = = = = = = = = = = = = = = = = = = = = = = = = = = = = = = =

\section{A Latent Patch Generative Model} %{Method}
\label{sec:method}

Our proposed approach,  which we refer to as {\sc LatentPatch},  enables the generation of novel images from a limited set of source images, and it consists of three steps, illustrated in Fig.~\ref{fig:schema}. Firstly, we construct a ``universal'' latent space. Secondly, we adapt this latent space to the source images. Finally, we generate new images in latent space and then decode them. We describe each of these steps in detail below.

\smallskip \noindent \textbf{Step 1. Construction of a ``universal'' representation space.} The objective of this step is to project the images into a generic, low-dimensional space with low spatial resolution, which does not depend on the source images. This makes it easier to capture the patch distributions of the source images. In our work, we use an off-the-shelf auto-encoder from VQ-GAN \cite{esser2021taming}, which has already been trained on a large generic dataset $\mathcal{D}$. An ideal network for this task should have sufficient capacity to compress any natural image at a low distortion rate, without significant overfitting between the reconstruction of test and train images. In practice, the encoder produces quantized images of size $M^2=$16×16 with $L=256$ channels.

% PCA
\smallskip \noindent \textbf{Step 2. Adapting the representation to the source images.} The manifold of source images $\mathcal{D}'$ only covers a small part of the representation space, which is meant to be universal.
\new{In order to significantly speed up our method, particularly the nearest neighbor patch search, we use a standard dimension reduction technique. In practice, we resort to a PCA whose parameters are precomputed over $\mathcal{D}'$ (corresponding to $P$ and $P^\star$ in Fig. \ref{fig:schema}). In this case, the dimensions are reduced from $L = 256$ to $r = 16$. This reduction has no noticeable impact regarding the quality of the reconstructions (\textit{cf.} Table \ref{tab:FID}), while making the patch search almost 16 times faster.}

\old{In order to make it easier for the generator 
to stay within the manifold of the source images, we aim to project the images into a space of even smaller dimension that is specialized for the source images.
We achieve this by applying a projection on the principal components of the latent code distribution.} %, inspired by~\cite{robb2020few}. pas sur que ce soit vraiment pareil
% \juju{je me suis trompé, ce sont les codes et non les patchs qui sont utilisé pour la PCA !  c'est avec l'algo de Raad que l'on calcule l'ACP des patchs ...}
%More precisely, let $z(x) \in \mathbb{R}^{L}$ denote a latent code at location $x\in \{0,..M-1\}^2$ in dimension $L$, given by the encoder.
%, and let a $\omega \times \omega$ patch be represented as a $\omega \times \omega \times L$ tensor defined as $p(x)[i,j] = \left[z(x - (i,j))\right]_{(i,j)\in{0,\omega-1}^2}$. 
%The reduced latent representation of a set of $B$ RGB images $U \in [0,1]^{B\times N\times N \times 3}$ is therefore given by $Z \in \mathbb{R}^{B\times M\times M \times L}$.

\old{To project the images into a smaller dimension, we first collect the codes at all location and batch indices to compute the ${L \times L}$ empirical covariance matrix $\Sigma$.}

%the matrix  $X \in \mathbb{R}^{(B M^2) \times L}$, % $X \in \mathbb{R}^{(B M^2) \times (L \omega^2)}$,
%as well as its mean $\bar X \in \mathbb{R}^{L}$. %$\bar X \in \mathbb{R}^{L\omega^2}$.
%We then compute the PCA of the empirical covariance matrix $\Sigma \in \mathbb{R}^{L \times L}$ % of the patch matrix:
% \begin{equation}
%     \label{eq:PCA_latent_patch}
%     \Sigma = \tfrac{1}{BM^2} (X - \un^T \bar X)^T (X - \un^T \bar X) = V^{T} D V
%     ,
% \end{equation}
%where $\mathbf{1}$ is a vector with all entries equal to one. 

\old{This matrix can be written as $\Sigma = V^{T} D V$ using principal component analysis.
% %Finally, 
The diagonal matrix $D$ is thresholded to keep the $r$ largest eigenvalues.
The projection $P$ in Fig.~\ref{fig:schema} is defined as 
$P : z(x) \rightarrow \sqrt{D'} V (z(x) - \bar X )$, 
%$P : p(x) \rightarrow \sqrt{D'} V (p(x) - \bar X )$, 
and the reconstruction as the operator 
$P^\star : \hat z (x) \rightarrow \sqrt{D'} V^{T} \hat z (x) + \bar X$.}

%\juju{Idee : comme il y a une quantification de l'espace latent, on pourrait précalculer les distances entre pixels et accélérer la recherche de plus proche voisin avec une look-up table ! Pourquoi ne l'a t'on pas fait ?}

\begin{figure}[tb]
    \centering
    \includegraphics[width = 1.0\linewidth]{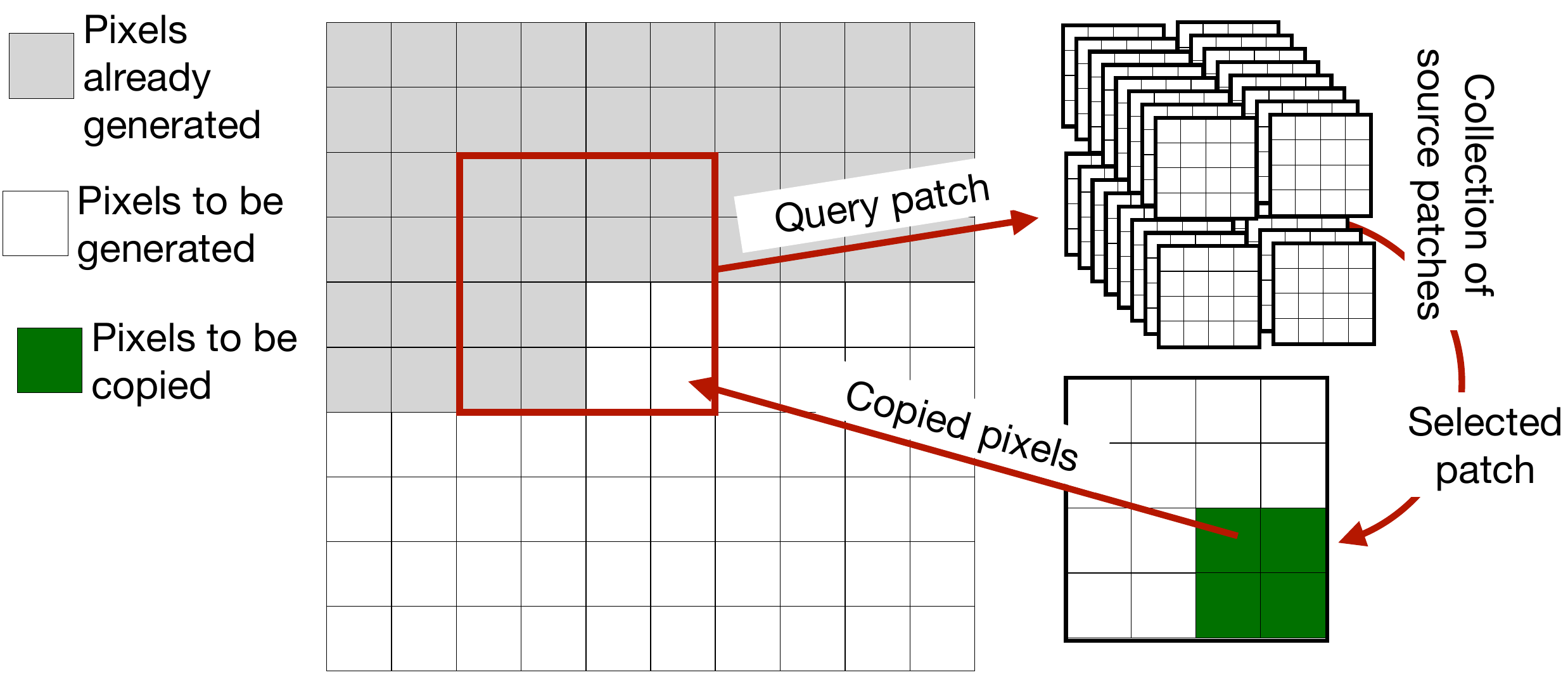}
    \caption{\small  Generation a $10\times10$ image using patches of size $4\times4$ and $2\times 2$ strides, following the same process at each scale. The query patches are masked to exclude the not-yet-generated pixels. 
    }
    \label{fig:patch_sampling}
    \vspace{-3.5em}
\end{figure}

\smallskip \noindent \textbf{Step 3. Generating images in the specialized representation space.}
As previously mentioned, our method is inspired by both texture synthesis and single image generative models, and focuses on the patch distributions of the source images. Unlike most existing approaches, we consider the joint latent patch distribution of the collection of source images.

More precisely, the generator, denoted as $G$, sequentially synthesizes latent codes $z(x)$, \new{at location $x\in \{0,..M-1\}^2$}, by sampling random codes from the empirical latent distribution, as  illustrated in Fig.~\ref{fig:schema}.
This non-parametric procedure does not require any training, contrary to other 
techniques based on transformers~\cite{esser2021taming}, auto-regressive synthesis~\cite{razavi2019generating,ben2022fewgan} or cross-attention \cite{rombach2022high}.
Likewise, it aims at predicting the value of a latent code $z(x)$ based on the observation of previously predicted values, in a (arbitrarily) raster-scan order.

In order to generate plausible combinations of latent codes, the proposed generator is based on patch sampling, similarly to nearest-neighbor patch-based texture synthesis \cite{efros1999texture, ashikhmin2001synthesizing}.
As shown in Fig.~\ref{fig:patch_sampling}, we consider here the latent $\omega \times \omega$ patch distribution of the source images $\mathcal{D}'$, rather than RGB patches.
Let a $\omega \times \omega$ patch at location $x$ be defined as a $\omega \times \omega \times L$ tensor defined as $p(x)[i,j] = \left[z(x - (i,j))\right]_{(i,j)\in\{0,\omega-1\}^2}$.
To improve the quality of the synthesis and increase the likelihood of the generated sequence, we make use of a stride $w$ when sampling query patches $p(x)$, as advocated in \cite{kwatra2005texture} for instance.
This means that only a small portion (of size $w \times w$) of the retrieved patch is actually copied, as the query is shifted by $w$ (from left to right).
%\juju{attention a differencier $s$ la taille du patch copiée et $\omega$ la taille du patch avec contexte dans le texte}

We adopt a multi-scale approach to impose long-range spatial correlations in the generated images. The generative model is initialized at the coarsest scale ($s=1$) by interpolating the $16 \times 16$ image given by the encoder $E$ to obtain a $10 \times 10$ spatial resolution. A random patch is then placed in the top-left corner. 
Query patches are masked at the coarsest scale to discard pixels that have not yet been generated (see Fig.~\ref{fig:patch_sampling}).
It is important to note that this mask is not necessary at larger scales or when using a reference image for initialization (as shown in Fig.~\ref{fig:teaser}~(d)), as all the pixels are available.
Once the synthesized image is upsampled to the next scale, it serves as a reference for generating the next level of detail, and this process continues until the desired $16 \times 16$ resolution is achieved at the finest resolution. The number of scale $S$ is a parameter of the algorithm.
Upsampling is performed using interpolation from the previous scale (see Fig.~\ref{fig:patch_sampling}).

% - k-NN patch sampling
We still have to specify how the patches of the source images are chosen for a given patch query.
Generating diverse images with limited data is a significant challenge, as it involves balancing the diversity of generated images with fidelity to the distribution of source images without overfitting. Positional embedding of the query features is a key aspect of training large generative models (see \emph{e.g.}~\cite{lin2021infinity,esser2021taming}). In this work, we sample patches $p(x)$ from the same location $x$ in example images to restrict the set of nearest-neighbor patches and speed-up the search. 
To ensure diversity, we uniformly sample the retrieved patch $p(x)$ from the $k>1$ nearest-neighbors, rather than copying the closest match. This approach, combined with a small stride $w$, prevents exact replication of the training data, as sampling neighboring patches from the same example image can be avoided by setting $k>1$.
This acts similarly as the temperature and top-k sampling parameters in likelihood-based models such as \cite{esser2021taming}.

%\juju{positiional embedding -> distribution conditionnelle selon x, citer toutes les méthodes qui sont similarire (comme infinity GAN)}
% diversity = small patch copy + random sampling from k-NN
%\juju{parler ici de l'algo de Raad ?}

% =======
% VQ-GAN trained on FFHQ+CelebAHQ
% VQ-GAN Latent space: 256×16×16
% PCA on the dataset to compress it to 16×16×16
% Compose new latent images based on L. Raad method
% Look for patches in the same position (faces are aligned)
% Let VQ-GAN decode the image
% Simpler approach compared to learning a transformer
% Can restrict the dataset -> few shot learning

%\lipsum[2]

\section{Experiments}
\label{sec:experiments}

\noindent\textbf{Experimental Settings.} For all experiments, we use an auto-encoder based on VQ-GAN~\cite{esser2021taming}, trained on the FFHQ~\cite{karras2019style} human face dataset, which contains a large number of high-quality images. FFHQ serves as the universal dataset $\mathcal{D}'$ for our experiments. The auto-encoder and quantizer have 72M parameters. 
The generative part of the method of \cite{esser2021taming}, is only used for comparisons.
Note that, as in~\cite{esser2021taming}, synthesized codes $z$ are encoded using the decompressed codebook before feeding the decoder. In all experiments, the sampling parameter is fixed to $k=3$, starting from $10\times 10$ images, up to $16\times 16$ over $S=5$ scales, except for referenced generation and edition which only require a single scale.

We conduct experiments using various sets of source images $\mathcal{D}'$, the images being randomly sampled from CelebA-HQ ~\cite{liu2015faceattributes}, with a resolution of $256$ pixels. 
To verify that the auto-encoder of~\cite{esser2021taming} is not overfitting, we use memorization detection techniques from \cite{webster2019detecting} and compare reconstruction errors of the source samples in $\mathcal{D}'$ with those from $\mathcal{D}$. Using the PCA, the original $M^2=$16×16 latent representation based on a $L=256$ dimensional codebook of $1024$ atoms is compressed to $r=16$ dimensions. We precompute the PCA over the latent features from the encoded source dataset $E(\cal{D}')$.

\begin{figure}[tb]
  \centering
  \setlength\tabcolsep{0.4mm}
  \setlength{\imgsize}{0.105\textwidth}
  \small
  \begin{tabular}{c ccccc}
  &\parbox{\imgsize}{\centering${\cal D'}$ index\\\scriptsize($B=16$)}
  & $B=16$ & $B=128$ & $B=1024$\\
  \rotatebox[origin=c]{90}{\parbox{\imgsize}{\centering\small $\omega=2$\\$w=1$}}
  &\includegraphics[valign=m,width=\imgsize]{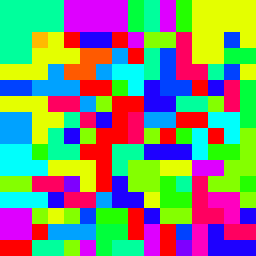}
  &\includegraphics[valign=m,width=\imgsize]{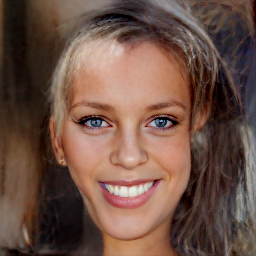}
  &\includegraphics[valign=m,width=\imgsize]{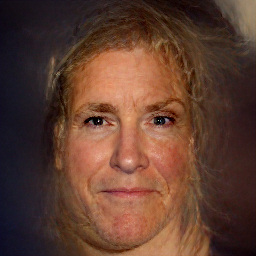}
  &\includegraphics[valign=m,width=\imgsize]{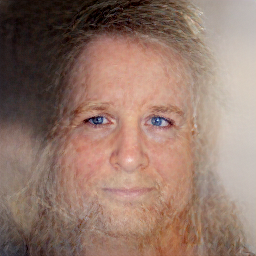}\\
  \rotatebox[origin=c]{90}{\parbox{\imgsize}{\centering\small $\omega=4$\\$w=2$}}
  &\includegraphics[valign=m,width=\imgsize]{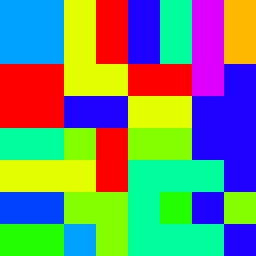}
  &\includegraphics[valign=m,width=\imgsize]{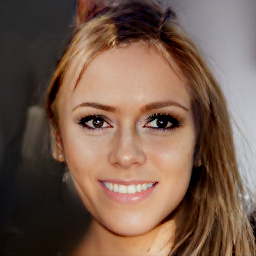}
  &\includegraphics[valign=m,width=\imgsize]{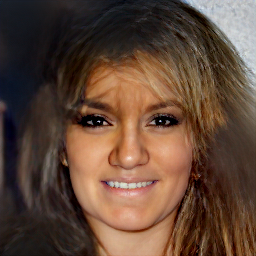}
  &\includegraphics[valign=m,width=\imgsize]{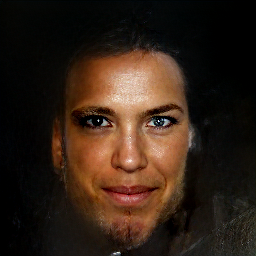}\\
  \rotatebox[origin=c]{90}{\parbox{\imgsize}{\centering\small  $\omega=6$\\$w=2$}}
  &\includegraphics[valign=m,width=\imgsize]{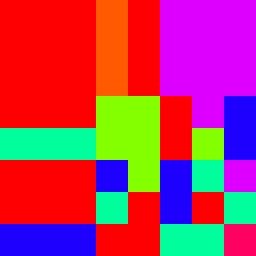}
  &\includegraphics[valign=m,width=\imgsize]{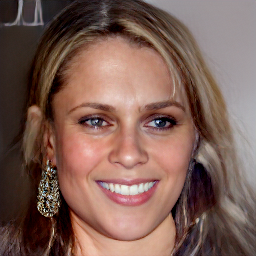}
  &\includegraphics[valign=m,width=\imgsize]{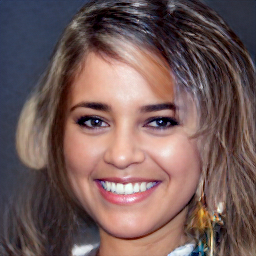}
  &\includegraphics[valign=m,width=\imgsize]{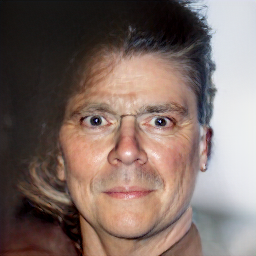}\\
  \rotatebox[origin=c]{90}{\parbox{\imgsize}{\centering\small $\omega=8$\\$w=3$}}
  &\includegraphics[valign=m,width=\imgsize]{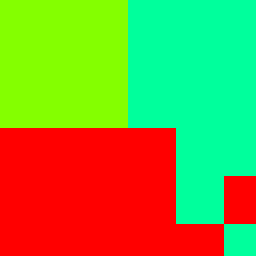}
  &\includegraphics[valign=m,width=\imgsize]{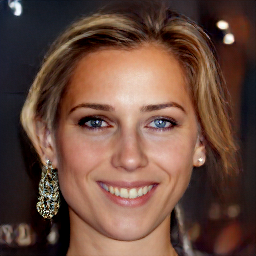}
  &\includegraphics[valign=m,width=\imgsize]{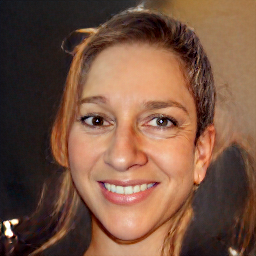}
  &\includegraphics[valign=m,width=\imgsize]{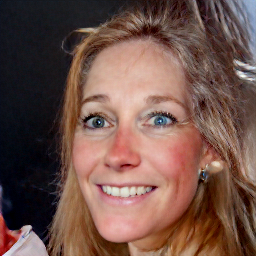}\\
  \end{tabular}
  \caption{\small
Comparison of generated images with different patch sizes ($\omega$), strides ($w$) and data size ($B$).
Coherence of the generated images improves with increasing patch size and stride, but larger example regions also result in reduced diversity. The first column shows the patch index, encoded in normalized hue values, for $B=16$. 
%A more complete set of results is available in~\cite{samuth2023webpage}.
}
  \label{fig:comparative_result}
  \vspace{-0.5em}
\end{figure}

\smallskip\noindent\textbf{Baseline.} Fig.~\ref{fig:teaser} (a \& b) shows that randomly copying codes $z(x)$ or $p(x)$ patches, even when sampling the same location $x$ than in example images, does not provide realistic faces. 
%In other words, the auto-encoder is not a projector that would generate images from any arrangement of latent codes. As expected, their implicit distribution is still highly relevant, justifying in that case the use of other methods such as {\sc LatentPatch}.
This shows that randomly sampling latent codes or patches alone is not sufficient and motivates the use of more sophisticated samplers such as the one proposed.

\smallskip\noindent\textbf{Random Face Generation.} The results shown in Fig.~\ref{fig:comparative_result} are obtained with our method by randomly sampling $B = |\cal D'|$ images from CelebA-HQ, and generating images with various patch sizes $\omega$ (with a fixed stride of $w = \lfloor \tfrac{1}{3}\omega +\tfrac{1}{2}\rfloor$). 
Computation time is fairly low for a sequential algorithm: after loading the model, a batch of 16 images is generated in 1 second for $B=16$. 
As expected, there is a trade-off between fidelity and diversity, where diversity is promoted by increasing $B$ and $k$. Indeed, doing so will allow the generative algorithm to pick more perceptually fitting patches from $\cal{D}'$. Decreasing both $w$ and $\omega$ will help generating more local variations, at the price of fidelity. Copying bigger patches taken from existing images implies that these are already locally coherent.
Note that the multi-scale scheme, which ensures coherent image generation, relies on latent image interpolation. This may create noticeable artifacts (first row of Fig.~\ref{fig:comparative_result}), due to the smoothing in the latent space.

\old{Finally, the point of the parameter $S$ is a simple way for our method to produce coherent outputs, as a $\omega \times \omega$ patch captures a wider portion of the image on the coarsest scales. This induces some color bias as shown on Fig. \ref{fig:comparative_result}, due to the bilinear downscaling artifacts in the latent space.}

\begin{table}[tb]
  \centering
  \iftrue % ajout colonne
    \small
    \begin{tabular}{c cc cc}
        %\hline
        %\multicolumn{3}{c}{CelebA-HQ $256 \times 256$}\\
        %\hline
        & Latent space & Method % \multicolumn{2}{c}{Method}                 
        & FID$\downarrow$ & \new{Diversity}$\uparrow$ 
        \\ 
        \hline
        \multirow{2}{*}{\rotatebox[origin=c]{90}{AE}}
        &VQ-GAN\textsubscript{F} & (Reconstruction)                          &  8.9 & 1.03
        \\
        &VQ-GAN\textsubscript{F} 
        & PCA                 &  8.9 & 1.03 
        \\
        \hline
        \multirow{4}{*}{\rotatebox[origin=c]{90}{Generation}}
        &VQ-GAN\textsubscript{C} 
        & Transformers\textsubscript{C} \cite{esser2021taming}   & 10.2 & 1.03 
        \\
        &{VQ-GAN\textsubscript{C}}
        & {\sc LatentPatch}\textsubscript{C}                                  & 31.6 & 0.80 
        \\
        &VQ-GAN\textsubscript{F}
        &{\sc LatentPatch}\textsubscript{C}                                  & 35.1 & 0.84 
        \\

        %\textbf{Ours->LatentPatch} (with attributes) & \ben{TODO}      & \ben{TODO} \\
        %\hline
        \cmidrule(lr){2-5}
        %&Baseline: &Random Patch Sampling                               
        & VQ-GAN\textsubscript{F} & \small Random sampling
        &123.0 & 0.85 \\ \hline
    \end{tabular}
  \else
    \begin{tabular}{ccc}
        %\hline
        %\multicolumn{3}{c}{CelebA-HQ $256 \times 256$}\\
        \hline
        \small Method                               & FID$\downarrow$ & Diversity$\uparrow$ \\ \hline
        VQ-GAN\textsubscript{F} Reconstruction                          &  8.9 & 1.03\\
        VQ-GAN\textsubscript{F} + PCA\textsubscript{16}                 &  8.9 & 1.03 \\
        \hline
        VQ-GAN\textsubscript{C} + Transformers \cite{esser2021taming}   & 10.2 & 1.03 \\
        \textbf{Ours}\textsubscript{F}                                  & 35.1 & 0.84 \\
        \textbf{Ours}\textsubscript{C}                                  & 31.6 & 0.80 \\
        %\textbf{Ours} (with attributes) & \ben{TODO}      & \ben{TODO} \\
        \hline
        {Baseline : Random sampling}                                    &123.0 & 0.85 \\ %\hline
    \end{tabular}
  \fi
  \caption{\small FID and diversity scores, both relative to $\cal{D'}=\text{CelebA-HQ}$, for the auto-encoder alone and as well as for various generative models. \new{For the FID, lower is better. For our diversity score, higher is better.}  Note that {\sc LatentPatch}\textsubscript{C} generates images using the latent space of the auto-encoder, either trained on F (FFHQ) or C (CelebA-HQ), but still use images from C as sources.
  %\juju{$E(C)\rightarrow P_C $ ?} \ben{Il faudrait définir la notation quelque part}
  }

 % Comparing VQ-GAN transformer-based generation fidelity and diversity with our approach. Random latent patch copy as shown in Fig.~\ref{fig:teaser}(b) is used as a baseline.
 % The auto-encoder (AE) is pre-trained over $\cal D$, that is, either the source dataset ($C$: CelebA-HQ) %\cite{liu2015faceattributes}) % j'ai enlevé les refs aux datasets bien connus pour gagner en place 
 % or an entirely different one ($F$: FFHQ \cite{karras2019style}). 
  %\juju{je ne suis pas sur de comprendre : c'est $D$ qui change et non $D'$ ? j'avais compris l'inverse : si c'est bien le cas, je mettrais plutot VQ-GAN\textsubscript{F} + Transformers\textsubscript{F}.}
  %\ben{Oui, c'est $D$ qui change, car l'AE est entraîné sur $D$. Cependant, pour que le FID ait du sens, il nous faut $D'$ = CelebA-HQ. Idéalement, il nous faudrait VQ-GAN\textsubscript{F} + Transformers\textsubscript{C}, mais cela requirerait un nouvel entraînement.}
 % For a fair comparison, we allow our method to pick patches from the entire dataset $\mathcal D'=C$ ($B=30$k), and set the nearest neighbor sampling to $k=10$ (instead of $3$ on the rest of the experiences).
 % }
  \label{tab:FID}\label{tab:perceptual_diversity}
  \vspace{-1em}
\end{table}

\smallskip\noindent\textbf{Quality and diversity assessment}. We evaluated the quality of the generated images by computing the FID score on 10k images. We introduce a new normalized score for diversity. It computes the ratio between the average perceptual distances between images, from respectively generated and source data.
In practice, we compute the average LPIPS \cite{zhang2018unreasonable} distances for 700 image pairs.
The FID and diversity scores are displayed in Table~\ref{tab:FID}.
The first two rows correspond to the auto-encoder alone, and show that the PCA has no significant impact on the quality of the reconstructed images. The next rows compare our method with \cite{esser2021taming}, which uses a transformer trained on large datasets, and we found that our {\sc LatentPatch} model achieves competitive results without the training \new{of an entire generator}, and with only a few source images. \new{Note that our diversity metric does not assess the visual quality of the generations, but only how close it is relative to CelebA-HQ.}
%The \textsc{LatentPatch}\textsubscript{F} model uses source images from FFHQ, while \textsc{LatentPatch}\textsubscript{C} uses source images from CelebA-HQ.
In these experiments, $w=6$, $\omega=2$, $S=1$ and $k=10$. For a fair comparison with \cite{esser2021taming}, {\sc LatentPatch} is able to pick patches from the entire dataset, thus $B = |\text{C}| = 30\text{k}$.
%\juju{faire si possible un calcul de FID en sélectionnant un sous-ensemble de D' avec des atttribus pour vérifier que la qualité augmente en prenant des images similaires}

%\begin{figure}[!htb]
%  \caption{\small
%    \ben{TODO: Generation from attributes + Face editing. Pick samples with the same attribute or edit it directly.}
%  }
%\end{figure}

\smallskip\noindent\textbf{Conditional image generation.} Our method allows for conditional generation of images, by conditioning on an input image which constitutes the first scale of the generation process. This means that our approach can regenerate faces that are similar to a reference image, using patches from the source images, as shown in Fig.~\ref{fig:teaser}~(d). Furthermore, by simply choosing random images with desired attributes (\emph{e.g.} from Celeb-A) in $\cal{D}'$, the generated images can exhibit the desired features, as illustrated in Fig.~\ref{fig:teaser}~(f). A perceptually homogeneous $\cal{D}'$ helps the generative model produce coherent outputs without requiring as many examples. However, not all attributes imply proximity between the images of the dataset (\emph{e.g.} ``hats''). In such cases, it is preferable to directly edit the image in the latent space of VQ-GAN, as in Fig.~\ref{fig:teaser}~(e), which is simple to do by copying the spatial vectors of the desired attribute over a reference image. The auto-encoder is powerful enough to blend the images together convincingly.

% =======
% Results (Copy vs. Gaussian model)
% With random patches
% VQ-GAN outputs

% Show source of patches

% About the quantization
% Evaluation (Diversity w/ FID)? Other datasets?

% Improvements

%\paragraph{}

%Additional experiments and details can be found on the project web page~\cite{samuth2023webpage}.

% pas tout à fait pret il me semble, a mettre sur la page web
%Figure~\ref{fig:patch-rgb} shows that even though aligned sampling is enough to generate seemingly believable faces, we still need the abstraction of an auto-encoder to blend properly the patches.
%
%Figure~\ref{fig:multi-scale} demonstrates the need for multi-scale synthesis to capture long-range correlations.
%
\iftrue 
\else
\begin{figure}[!htb]
    \centering
    \begin{subfigure}[h]{\linewidth}
        \centering
        \setlength{\imgsize}{0.3\textwidth}
        
        \caption{\footnotesize Patch sampling in image space (RGB) \juju{with 1 scale ?}}
        \begin{tabular}{ccc}
            %\raisebox{.3\imgsize}{\rotatebox[origin=c]{90}{\parbox{\imgsize}{\centering\small Patch size $\omega$}}}&
            \includegraphics[width=\imgsize]{pictures/experiments/ablation/w=16.png}    &\includegraphics[width=\imgsize]{pictures/experiments/ablation/w=64.png} &\includegraphics[width=\imgsize]{pictures/experiments/ablation/w=96.png}\\
            $\omega=16$ & $\omega=64$ & $\omega=96$
        \end{tabular}
        \label{fig:patch-rgb}
    \end{subfigure}\\[-2mm]%
    %\vspace*{2em}
    \begin{subfigure}[h]{\linewidth}
        \centering
        \setlength{\imgsize}{0.3\textwidth}
        
        \caption{\footnotesize Multi-scale \juju{in latent space}}
        \begin{tabular}{ccc}
            %\raisebox{.3\imgsize}{\rotatebox[origin=c]{90}{\parbox{\imgsize}{\centering\small \#Scale $S$}}}&
            \includegraphics[width=\imgsize]{pictures/experiments/ablation/S=1.png}    &\includegraphics[width=\imgsize]{pictures/experiments/ablation/S=3.png} &\includegraphics[width=\imgsize]{pictures/experiments/ablation/S=5.png}\\
            $S=1$ & $S=3$ & $S=5$
        \end{tabular}
        \label{fig:multi-scale}
    \end{subfigure}
    \vspace*{-1mm}
    \caption{\small Ablation study. %(\ref{fig:patch-rgb}) shows that even though aligned sampling is enough to generate seemingly believable faces, we still need the abstraction of an auto-encoder to blend properly the patches.
    See the text for details.}
\end{figure}
\fi
%\vspace{-1em}
\section{Conclusion and future work}
\vspace{-0.5em}
This work proposes {\sc LatentPatch}, a simple non-parametric model for generating near photo-realistic face images from tiny datasets, using a coarse-to-fine patch sampling approach. The model has several advantages, including not requiring \new{the training of a generator} and being versatile enough for related tasks like image editing and conditional generation.  Future work will explore the possibility of generation from non-registered images, and the use of generic and lightweight auto-encoder.

\thanks{\footnotesize \textbf{Acknowledgment.} This work is partially supported by the project ANR-19-CHIA-0017.}
\bibliographystyle{IEEEbib}

{\small
    \bibliography{refs}
}

\end{document}